
\magnification=1200
\baselineskip=20pt
\def\lsim{<\kern-2.5ex\lower0.85ex\hbox{$\sim$}\ }
\def\rsim{>\kern-2.5ex\lower0.85ex\hbox{$\sim$}\ }

\overfullrule=0pt
\centerline{\bf GRAVITATIONAL PHASE OPERATOR AND COSMIC
STRINGS}
\vskip .5cm
\centerline{J. Anandan}
\centerline{Department of Theoretical Physics}
\centerline{University of Oxford}
\centerline{1 Keble Road}
\centerline{Oxford OX1 3NP, U. K.}
\centerline{and}
\centerline{Department of Physics and Astronomy}
\centerline{University of South Carolina}
\centerline{Columbia, SC 29208, USA}
\vskip .5cm
\vskip .5cm
\centerline{\bf \underbar{Abstract}}
A quantum equivalence principle is formulated by means of a
gravitational phase operator which is an element of the Poincare
group.
This is applied to the spinning cosmic string which suggests that it
may (but not necessarily)
contain gravitational torsion. A new exact solution of the Einstein-
Cartan-Sciama-Kibble equations for the gravitational field with
torsion is obtained everywhere for a cosmic string with uniform
energy
density,
spin density and flux. A novel effect due to the quantized
gravitational field of the cosmic string on the wave function of a
particle outside the string is used to argue that
spacetime points are not meaningful in quantum gravity.
\vskip .5cm
\vskip .5cm
\vskip .5cm
\vskip .5cm
\vskip .5cm
\vskip .5cm
To be published in Physical Review D. GR-QC/9507049.
\vfil\eject

\noindent{0. INTRODUCTION: RELATIVIZING AND QUANTIZING
GRAVITY}

After the discovery of special relativity by Lorentz, Poincare, and
Einstein, there was the problem of ``relativizing gravity'',
analogous
to the problem of ``quantizing gravity'' which exists today. It was
clear that Newtonian gravity was incompatible with special
relativity and it was necessary to replace it with a relativistic
theory of
gravity.
While several attempts were made to
do this, Einstein succeeded in constructing such a theory because
he
used i) the {\it
geometrical} reformulation of {\it special relativity} by
Minkowski, and
ii) the {\it operational}
approach of asking what may be learned
by probing gravity using {\it classical}
particles.

An important ingredient in (i) was Einstein's realization that the
times in the different inertial
frames, $t$ and $t'$, in the Lorentz transformation were on the
same footing. I.e. the {\it
interpretation} Einstein gave to special relativity, whose basic
equations were already known
to Lorentz and Poincare, was crucial to the subsequent work of
Minkwoski. It enabled
Einstein to get rid of the three dimensional ether, and
thereby pave the way for the
introduction of the four dimensional `ether', called space-time, by
Minkowski. By means of
(ii), Einstein concluded that the aspect of Newtonian gravity which
should be retained when this theory is modified is the
equivalence principle. This principle is compatible with special
relativity locally. This may be seen from the physical formulation
of the strong equivalence principle according to which in the
Einstein elevator that is freely falling in a gravitational field the
laws of special relativity are approximately valid. But this
principle allowed for the
modification of special relativity to incorporate gravity as
curvature of space-time.

Today we find that general relativity, the beautiful theory of
gravity which Einstein discovered in this way, is incompatible
with quantum theory. Can we then adopt a similar approach? This
would mean that we
should use 1) a {\it geometrical} reformulation of {\it quantum
theory}, and 2) an {\it
operational} approach of asking what may be learned
by probing gravity using {\it quantum}
particles.

As for (1), the possibility of using group elements as 'distances' in
quantum theory, analogous to space-time distances in classical
physics, was studied previously [1]. For a particular quantum
system, the corresponding representations of these group
elements may be used to relate points of the projective
Hilbert space, i.e. the set of
rays of the Hilbert space, which is the quantum generalization of
the classical phase space [2]. Recent
work on protective observation of the quantum state has shown
that the points of the projective
Hilbert space
are real, in the sense that they could be observed by
measurements on an individual system,
instead of using an ensemble of identical systems [3].

As for (2), the question is whether the motion of a quantum
system in a gravitational
field enables us to identify the aspect of general relativity
which must be preserved when this theory is replaced by a
quantum theory of gravity, i.e. the quantum analog of the
equivalence principle. In section 1,
I shall formulate such a principle. This will be applied to
cosmic strings, in section 2, because of their interesting
topological, geometrical, and quantum gravitational aspects. I shall
present an exact solution
of the Einstein-Cartan-Sciama-Kibble gravitational field equations,
valid in the interior as well
as the exterior of the cosmic string, which depends on three
parameters.

It will be shown in section 3 that when the
gravitational field of the string is quantized so that different
geometries may be superposed, the wave function of a test
particle even in a simply connected region is affected although
each of the superposed geometries is flat in this region. But a
special case of this
effect is invariant under a quantum diffeomorphism that
transforms different geometries differently, as discussed in
section 4. This freedom suggests that the points
of space-time have no
invariant meaning. So, there seems to be a
need to get rid of the four dimensional `ether',
namely space-time, in order to
incorporate the quantum diffeomorphism symmetry into
quantum gravity.
\vskip .5cm
\noindent{1. THE EQUIVALENCE PRINCIPLE IN CLASSICAL AND
QUANTUM PHYSICS}

First, consider the classical weak equivalence principle (WEP), due
to Galileo and
Einstein.
This has two aspects to it: In a space-time manifold with a
pure
gravitational field, a) the possible motions of all freely falling test
particles are the same, and b) at any point $p$ in space-time,
there
exists a neighborhood $U(p)$ of $p$ and a coordinate system
$\{x^\mu , \mu=0,1,2,3\}$ in this neighborhood such that the
trajectory of {\it every}
freely falling test particle through $p$ satisfies [4]
$${d^2x^\mu\over d\lambda^2}=0, \eqno(1.1)$$
at $p$ for a suitable parameter $\lambda$ along the trajectory.
This is
the
local form of the law of inertia and
the above coordinate system is said to be locally inertial at $p$.
The
condition (b) is a special property of the gravitational field, not
shared by any other field. For example, in an electromagnetic field
test particles with the same charge to mass ratio would satisfy
(a) but not (b). (The Lorentz 4-force is
proportional to the electromagnetic field strength which, being a
tensor, cannot be coordinate transformed away unlike the
connection
coefficients.)

Using (b), for massive and massless particles, it may be shown
that there exists an affine connection
$\omega$ such that the trajectories of freely falling test particles
are
affinely parametrized geodesics with respect to it [4]. Suppose
$\epsilon
= {d\over L}$, where $d\sim$ linear dimensions of $U(p)$ and
$L\sim$ radius of curvature obtained from the curvature
components of this connection, and we can neglect second orders
in
$\epsilon$. Such a neighborhood will be called a first order
infinitesimal neighborhood of $p$, and denoted by $U_\epsilon
(p)$.
Using the geodesic deviation equation, it may be shown that the
velocities of the freely falling test particles in $U_\epsilon(p)$ are
constant in an appropriately chosen coordinate system. This is a
stronger form of the WEP than its usual statement given above,
and
will be called the modified classical weak equivalence principle. It
is
valid in Newtonian gravity as well as Einsteinian gravity.

The above formulations of WEPs may be stated using
only
an affine connection and do not require a metric. In
$U_\epsilon$, the
affine structure defined by this connection has as its symmetry
group the
affine group $A(4)$ that is generated by the general linear
transformations and translations in a $4$ dimensional real vector
space. In the non relativistic limit, as the null cones `flatten',
$A(4)$ remains the symmetry group. In classical physics,
the interactions between the particles
restrict the symmetry group in $U_\epsilon$ to the
inhomogeneous
Galilei group (non relativistic physics), or the Poincare group $P$
(relativistic physics), which are both subgroups of $A(4)$. The
existence of this residual symmetry group in $U_\epsilon$ is a
form of the classical strong equivalence principle (SEP) valid for
relativistic and non relativistic gravity. In this way, non flat
space-time geometry may also in
some sense be brought into the frame-work of Felix Klein's
Erlanger program according to which a
geometry
is determined as the set of properties invariant under a symmetry
group [1].

What fundamental aspects about the gravitational field may be
learned if it is probed with quantum particles, instead of with
classical particles as in the above treatment? It was shown that
the
evolution of a freely falling wave function is given, in the WKB
approximation, by the action on the initial wave function by the
operator [5]
$$\Phi_\gamma =  Pexp[- i\int_{\gamma} \Gamma_\mu dx^\mu
],\eqno(1.2)$$
where
$$\Gamma_\mu ={\theta_\mu}^a
P_a +{1\over2}{{{\omega}_\mu}^a}_b {M^b}_a . \eqno(1.3)$$
which will be called the gravitational phase operator. Here the
energy-momentum operators $P_a$ and the angular
momentum operators ${M^b}_a, a,b = 0,1,2,3$ generate the
covering
group of the Poincare group $\tilde P$ that is a semi-direct
product
of $SL(2,C)$ and space-time translations $R(4)$. The fact that
mass
$m$ is a good quantum number in curved space-time and $m^2$
is
a
Casimir operator of $P$ already suggests that $P$ is relevant in
the
presence of gravity.

For every space-time point $p$, let $H_\epsilon (p)$ be the Hilbert
space of wave functions in $U_\epsilon (p)$ in which $\tilde P$
acts.
Owing to the linearity of the action of (1.2), it determines also the
evolution of any freely falling wave packet which can be
expanded
as a linear combination of WKB wave functions, provided the size
of
the wave packet is small compared to the radius of curvature, i. e.
it
is contained primarily inside $U_\epsilon$ at each point along
$\gamma$ which may be chosen to be along the center of the
wave
packet. This will be called the quantum weak equivalence
principle,
because (1.2) is a Poincare group element
independent of the freely falling
wave packet. In this respect, it is like the classical WEP according
to which the affine connection determined is independent of the
test
particle used.

In quantum physics, because the wave packet must necessarily
have
some spread, the WEP cannot be formulated by particle
trajectories
as in conditions (a) and (b) above, and it is necessary to use at
least
the neighborhood $U_\epsilon$. Indeed (1.2) was obtained [5]
using
the Klein-Gordon [6] and Dirac equations [7] which are covariant
under
$\tilde P$ in $U_\epsilon$. So, in quantum physics there is a close
connection between the quantum WEP, as formulated above, and
the quantum SEP
according
to which $\tilde P$ is the symmetry group of all laws of physics in
$U_\epsilon$. It is well known that (a) cannot be valid in
quantum physics, because the motions of wave functions depend
on their masses [8].
But the modified classical WEP and the classical SEP as stated
above
have the advantage that they have a smooth transition to
quantum
physics.

The above approximate concepts may be made mathematically
precise as follows: Each neighborhood $U_\epsilon (p)$ may be
identified with the tangent space at $p$ regarded as an affine
space.
The motions of freely falling test particles relate affine spaces
associated with two neighboring points by a linear transformation
and a translation, generated by $P_a$.
This gives a natural connection on the affine bundle[9] over
spacetime which is a principal fiber bundle with $A(4)$ as the
structure group. This is the connection used above to express the
modified
classical WEP. The quantum WEP requires the Poincare subbundle
with $\tilde P$ (to admit Fermions) as the structure group. Then
(1.3) defines a connection in
this
principal fiber bundle. The gravitational phase operator (1.2)
parallel transports with respect to this connection along the curve
$\gamma$. The above Hilbert space bundle, that is the union of
$H_\epsilon (p)$ for all space-time points $p$, is a vector
bundle associated to this principal fiber bundle with a connection
that is the representation of (1.3) in this Hilbert space.

The curvature of the above connection is the Poincare Lie algebra
valued 2-form
$$F =d\Gamma+\Gamma\wedge\Gamma =
Q^aP_a +{1\over2}{R^a}_b {M^b}_a ,\eqno(1.4)$$
where, on using (1.3) and the Lie algebra of the Poincare group,
$$Q^a=d\theta^a + {\omega^a}_b\wedge\theta^b ,{R^a}_b
=d{\omega^a}_b+{\omega^a}_c\wedge{\omega^c}_b .\eqno(1.5)$$
which are called respectively the
torsion and the linear curvature.
If the wave equation used to obtain (1.2) did not contain torsion,
then the torsion in (1.4), of course, is also zero. However, the
above modified classical WEP and the quantum WEP make it
natural
to have torsion and suggest that if the torsion is zero then there
should be a
good physical reason for it.

Suppose $\gamma$ is a closed curve. Then (1.2) is a holonomy
transformation determined by the above affine connection.
It may then be
transformed to an appropriate integral over a 2-surface $\Sigma$
spanned by $\gamma$ as follows. Let $O$ be a fixed point in
$\Sigma$. Foliate $\Sigma$ by a 1-parameter family of curves
$\lambda (s,t)$, where $s\epsilon [0,1]$ labels the curves and
$t\epsilon [0,1]$ is the parameter along each curve. All curves
originate at $O$, which corresponds to $t=0$. The $(s,t)$
are smooth coordinates on $\Sigma$ excluding $O$. Suppose
$\gamma$ begins and ends at
$(0,1)$. Let
$\Lambda (s,t)={\Phi}_{\hat\lambda (s,t)}$, where $\hat\lambda
(s,t)$ is the segment of a curve of the above family that begins at
$O$ and ends at $(s,t)$. Then
$$\Lambda^{-1} (0,1)\Phi_\gamma \Lambda (0,1) =
P_{st}exp[- i\int_0^1 ds\int_0^1 dt \Lambda^{-1} (s,t)
F_{\mu \nu}(s,t)\Lambda (s,t)\ell^\mu m^\nu ],\eqno(1.6)$$
where $\ell^\mu ={\partial x^\mu\over\partial s},
m^\mu ={\partial x^\mu\over\partial t}$
and $P_{st}$ means surface ordering, i. e.
in
the expansion of (1.6) terms with greater value of $s$ precede
terms
with smaller value of $s$, and for equal values of $s$ terms with
greater value of $t$ precede terms with smaller value of $t$. In
(1.6) all field variables are transported
to the common point $O$ so that the integrals are meaningfully
performed in the affine space at $O$.

To prove (1.6), note that the LHS of (1.6) is a holonomy
transformation which begins and ends at $O$, and may be
expressed
as a product of holonomy transformations $\Phi_s$ over triangles
whose sides are two $s=constant$ curves and an infinitesimal
segment of
$\gamma$. Each $\Phi_s$ may be written as a product of
$\Phi_{st}$
over infinitesimal ``rectangles'', bounded by $s=$ const. , $t=$
const.
curves, which are transported to $O$, which yields (1.6). This
extends a
known result for Yang-Mills field and linear curvature [10] to
include torsion.

It follows from (1.6) that in the absence of gravity in a simply
connected region (1.2) is path independent. I shall take the
equivalent
statement that  the path dependence of (1.2) implies gravity as
the
{\it definition} of the gravitational field
even when the region is not
simply connected. This definition makes the converse of this
statement also valid. So, by probing
gravity using quantum mechanical systems, without
paying any attention to gauge fields, gravity may be
obtained naturally as a
Poincare gauge field in the sense
of Yang's integral formulation of gauge field [11].

An advantage of this point of view is that it also provides a
unified description of gravity and gauge fields. If a wave function
is interacting not only with the gravitational field but also other
gauge fields, then its propagation in the WKB approximation is
given by the action of an operator of the form (1.2) with
$$\Gamma_\mu ={\theta_\mu}^a
P_a +{1\over2}{{{\omega}_\mu}^a}_b {M^b}_a +{A_\mu}^jT_j ,
\eqno(1.7)$$
where ${A_\mu}^j$ is the Yang-Mills vector potential and $T_j$
generate the gauge group $G$. So, (1.2) now is an element of the
entire symmetry group, namely $\tilde P\times G$. Thus, unlike
the
classical WEP, the quantum WEP naturally extends to incorporate
all gauge fields.

The above fact that the observation of all the fundamental
interactions
in nature is via elements of the symmetry group suggest a
symmetry ontology. By this I mean that the elements of
symmetry group are observable and therefore real. Moreover, the
observables such as energy, momentum, angular momentum and
charge, which are usually observed in quantum theory are some
of the generators of the above symmetry group. Observation
always requires interaction between the observed system and the
apparatus. Ultimately, these interactions are mediated by gravity
and gauge fields. I therefore postulate that the only observables
which can actually be observed are formed from the generators of
symmetry group, which according to our current understanding of
physics are generators of $\tilde P\times G$. Symmetry is destiny.

\vskip .5cm
\noindent{2. COSMIC STRING - AN EXACT SOLUTION}

As an example, consider cosmic strings, which are predicted by
gauge
theories [12] and are of astrophysical interest because of their
possible role in galaxy formation [13] and as gravitational lenses
[14,15]. Consider a cosmic string whose axis is along the $z-$axis.
Since the torsion and curvature outside the string are zero, its
exterior geometry is determined entirely by the affine holonomy
transformation associated with a closed curve $\gamma$ going
around the
string, given by (1.2). But owing to the cylindrical symmetry of
this geometry,
this transformation should commute with ${M^2}_1$ which
generates rotations about the axis of the string.
The most general affine holonomy transformation that is
restricted to the Poincare group by (1.3) which
commutes with
${M^2}_1$ is of the form
$$\Phi_\gamma= \exp[-i  (b P_o + c P_3 +  a {M^2}_1 + d {M^3}_0)]
.\eqno(2.1)$$
Therefore, the most general external geometry should depend on
the four
parameters $a, b, c$ and $d$. This geometry has been obtained,
from the
point of view of affine holonomy, by Tod [16] although the
present
argument which uses the gravitational phase operator (1.2) is
somewhat simpler and more physical.

I shall consider here only the most general {\it stationary}
exterior
solution which depends only on three parameters ($d=0$):
$$ds^2 = (dt + \beta d\phi )^2 -
d\rho^2-\alpha^2 \rho^2 d\phi^2-(dz+\gamma d\phi )^2
,\eqno(2.2)$$
where $\alpha,\beta$ and $\gamma$ are constants related to $a,
b$ and $c$ respectively. The external metric (2.2) was due to
Gal'tsov and Letelier [17]. The special
case of
$\gamma =0$ was previously considered by Deser et al [18] and
Mazur [19].
It is worth noting that the usual linear holonomy around the
cosmic string can only determine the parameter $\alpha$.
Whereas the translational part of the affine holonomy
distinguishes metrics (2.2) with different values for ($\beta,
\gamma$) [16], which shows the importance of affine holonomy
even in this purely classical context.
It follows from (1.4) and (1.5) that the rotational
part of the affine
holonomy,
due to $\alpha$, requires curvature inside the string.
The translational part of the affine holonomy, due to $\beta$ and
$\gamma$, suggests (but does not require) the inclusion of torsion
inside the string.

With a view towards this, rewrite (2.2) as
$ds^2=\eta_{ab}\theta^a \theta^b$,
where the orthonormal co-frame field $\theta^a$ is
$$\theta^0= dt+\beta d\phi , \theta^1 =d\rho , \theta^2
=\alpha\rho
d\phi , \theta^3=dz+\gamma d\phi
.\eqno(2.3)$$
Let $e_a$ be the frame (vierbein) dual to $\theta^a$:
${\theta_\mu}^b{e^\mu}_a=\delta_a^b$. The connection
coefficients in this basis are
${{\omega_\mu}^a}_b \equiv {\theta_\nu}^a\nabla_\mu
{e^\nu}_b=0$,
for all $a,b,\mu$ except for
$${{\omega_{\hat \phi}}^1}_2=-{{\omega_{\hat \phi}}^2}_1 =-
\alpha , \eqno(2.4)$$
assuming no torsion in the exterior.
This external geometry is {\it affine flat}, i.e. $Q^a=0,{R^a}_b=0$ on
using (1.5), and yet the
{\it affine holonomy around the string is non trivial} [20].

Suppose $\gamma$ is a closed curve around the string. Then from
(1.2),
$$\eqalignno{\Phi_\gamma &=  exp\left(- i\oint_{\gamma}
{\theta_\mu}^0
P_0 dx^\mu \right) exp\left(- i\oint_{\gamma} {\theta_\mu}^3
P_3 dx^\mu \right)\cr
&\times Pexp\left[- i\oint_{\gamma}
\left(\sum_{k=1}^2{\theta_\mu}^k
P_k+{{{\omega}_\mu} ^1}_2 {M^2}_1\right)dx^\mu
\right]. &(2.5)\cr}$$
The three factors in (2.5) commute with one another. On
comparing
with (2.1) and using (2.3), $b=2\pi \beta$ and $c=2\pi \gamma$.
The
first factor in (2.5), which is a time
translation, may be given a physical meaning as follows: Suppose
an optical, neutron or superconducting interferometer encloses
the string once and is at rest with respect to the above coordinate
system. Then the above time translation gives rise to a ``Sagnac''
phase shift [6,21,19,22], which in the present case is $\Delta\phi_E
=
2\pi\beta E$, where $E$ is the frequency of the interfering
particle (eigenvalue of $P_0$).

The second factor in (2.5), which is a spatial translation, may be
given physical meaning by the following new effect: Suppose the
beam at the beam splitter of the above
interferometer has a $z-$
component of momentum
$p$. I.e. $p$ is the approximate eigenvalue of $P_3$. Then, this
factor gives rise to the phase shift $\Delta\phi_p =
2\pi\gamma p$.

If in (1.6), coordinates and basis can be chosen such that
$\Lambda (s,t)\simeq 1$ for all $s,t$, then $\Sigma$ will be called
infinitesimal. It follows from (1.6), (1.4) and (2.5) that when the
cross-
section
of the string is infinitesimal in this sense, it must necessarily
contain
torsion in order that the surface integral has the time translation
contained in the line integral. Then $\Delta\phi$ may be regarded
as a topological phase shift due to the
enclosed torsion inside the string. It is possible for the string not
to
contain torsion, but only by violating the above
infinitesimality assumption.

The simplest gravitational field
equations in the presence of torsion are the Einstein-Cartan-
Sciama-
Kibble (ECSK) equations [23], which may be written in the form
[24]
$${1\over 2}\eta_{ijkl}\theta^l\wedge R^{jk}=-8\pi G t_i
,\eqno(2.6)$$
$$\eta_{ijkl}\theta^l\wedge Q^k = 8\pi G s_{ij} ,\eqno(2.7)$$
where $t_i$ and $s_{ij}$ are 3-form fields representing the
energy-momentum and spin densities. I shall now obtain an exact
solution of these equations for the interior of the cosmic string
which
matches the exterior solution (2.2). This will then give physical
and geometrical meaning to the parameters $\alpha$, $\beta$ and
$\gamma$ in (2.2). This
solution will be different from earlier torsion string solutions [25]
which have static interior metrics matched with exterior metrics
which are different from (2.2).

The $\rho$ and $z$ coordinates in the interior will be chosen to be
the distances measured by the metric in these directions. Since
the
exterior solution has symmetries in the $t, \phi$, and $z$
directions,
it is reasonable to suppose the same for the interior solution. So,
all functions in the interior will be functions of
$\rho$ only. So, I make the following ansatz
in
the interior:
$$\theta^0 =u(\rho) dt + v(\rho) d\phi , \theta^1 = d\rho , \theta^2
=
f(\rho) d\phi , \theta^3 = dz + g(\rho )d\phi , {{\omega}
^2}_1=k(\rho) d\phi =
-{{\omega} ^1}_2 ,\eqno(2.8)$$
all other components of ${{\omega} ^a}_b$ being zero, and
$ds^2=\eta_{ab}\theta^a\theta^b \equiv g_{\mu\nu}dx^\mu
dx^\nu$. Suppose also that there is a fluid in the interior whose
energy density $\epsilon$ and spin
density $\sigma$ polarized in the $z$-direction are constant, and
this spin has a constant current density $\tau$ in the z-direction.
I. e.
$$\eqalignno{t_0 &= \epsilon \theta^1\wedge \theta^2 \wedge
\theta^3 = \epsilon
f(\rho)d\rho\wedge d\phi \wedge dz ,\cr
s_{12} =-s_{21}&=\sigma
\theta^1\wedge \theta^2 \wedge \theta^3 -\tau \theta^0\wedge
\theta^1 \wedge \theta^2 \cr
&= \sigma
f(\rho)d\rho\wedge d\phi \wedge dz-\tau uf(\rho) dt\wedge
d\rho \wedge d\phi ,  &(2.9)\cr}$$
the other components of $s_{ij}$ being zero.
In terms of the components of the energy-momentum and spin
tensors in the present basis, this means that ${t^0}_0 =\epsilon=$
constant and ${s^0}_{12} = \sigma =$ constant.

 The torsion and curvature components here are defined by $Q^k =
Q_{\mu\nu}{}^k dx^\mu\wedge dx^\nu$ and $R^{jk}
=R_{\mu\nu}{}^{jk} dx^\mu\wedge dx^\nu$.
It is assumed that there is no surface energy-momentum or spin
for
the string. Then the metric must satisfy the junction conditions
[26], which in the present case are
$$g_{\mu\nu}|_-=g_{\mu\nu}|_+ , \partial_{\hat\rho}
g_{\mu\nu}|_+
=\partial_{\hat\rho} g_{\mu\nu}|_- + 2K_{(\mu\nu)\hat\rho} ,
\eqno(2.10) $$
where
$K_{\alpha\beta\gamma} = {1\over 2}(-
Q_{\alpha\beta\gamma}+Q_{\beta\gamma\alpha}-
Q_{\gamma\alpha\beta})$ is the contorsion or the defect tensor,
$|_+$ and $|_-$ refer to the limiting values as the boundary of the
string is approached from outside and inside the string,
respectively,
and the hat denotes the corresponding coordinate component.

Substitute (2.8), (2.9) into the Cartan equations (2.7). The
$(i,j)=(0,2),(0,3),(2,3)$ eqs. are automatically satisfied.
The $(i,j)=(0,1),(1,3),(1,2)$ eqs. yield
$$f'(\rho)=k(\rho), u'(\rho) = 0 , v'(\rho) = 8\pi G\sigma
f(\rho), g'(\rho) = 8\pi G\tau
f(\rho), \eqno(2.11)$$
where the prime denotes differentiation with respect to $\rho$.
Therefore, the continuity of the metric (eq. (2.10)) implies that
since $u=1$ at the boundary, $u(\rho)=1$ everywhere. Now
substitute (2.8), (2.9) into the Einstein equations (2.6). The $i=0$
eq. yields
$$k'(\rho) = -8\pi G\epsilon f(\rho) . \eqno(2.12)$$
The $i=1,2,3$ equations yield, respectively
$$t_1=0,t_2=0,t_3
= {k'\over 8\pi G}dt\wedge d\rho\wedge d\phi
=-\epsilon\theta^0\wedge\theta^1\wedge\theta^2,\eqno(2.13)$$
using (2.12). Hence, ${t^3}_3 = \epsilon = {t^0}_0$. From (2.11) and
(2.12),
$$f''(\rho) + {1\over {\rho *}^2}f(\rho) = 0 , \eqno(2.14) $$
where $\rho * =(8\pi G\epsilon )^{-1/2}$.
In order for there not to be a metrical ``cone'' singularity at $\rho
= 0$, it is necessary that
$\theta^2\sim \rho d\phi$ near $\rho =0$.
Hence, the solution of (2.14) is $f(\rho)
= \rho* sin{\rho\over \rho *}. $
Then from (2.11), $k(\rho)=cos{\rho\over \rho *}$, and requiring
$v(0)=0=g(0)$ to avoid a conical
singularity, $v(\rho) = 8\pi G\sigma \rho *^2\left(1-
cos{\rho\over \rho *}\right), $ and $g(\rho) = 8\pi G\tau \rho
*^2\left(1-
cos{\rho\over \rho *}\right). $

This gives the metric in the interior of the string to be
$$\eqalignno{ds^2 &= \left[ dt + 8\pi G\sigma \rho *^2\left(1-
cos{\rho\over
\rho*}\right) d\phi \right]^2
-d\rho^2 -\rho*^2 sin^2\left({\rho\over \rho *}\right) d\phi^2 \cr
&-\left[ dz+8\pi G\tau \rho *^2\left(1- cos{\rho\over
\rho*}\right) d\phi \right]^2,
&(2.15)\cr} $$
and the connection is ${{\omega} ^2}_1=cos{\rho\over
\rho*} d\phi $. The only non vanishing components of torsion and
curvature in the interior are
$$\eqalignno{Q^0 &= 8\pi G\sigma\rho *
sin\left({\rho\over \rho *}\right) d\rho\wedge
d\phi , Q^3 = 8\pi G\tau \rho *
sin\left({\rho\over \rho *}\right) d\rho\wedge
d\phi ,\cr {R^1}_2&={1\over\rho*}sin\left({\rho\over \rho
*}\right)
d\rho\wedge
d\phi=-{R^2}_1.&(2.16)\cr}$$

I apply now the junction conditions (2.10), which will
show that $\rho$ is discontinuous across the boundary. Denote the
values of
$\rho$ for the boundary in the internal and external coordinate
systems by $\rho_-$ and $\rho_+$ respectively.
 From (2.1) and (2.15), $g_{\hat t\hat \phi}$,$g_{\hat z\hat \phi}$,
and $g_{\hat \phi\hat
\phi}$
are respectively continuous iff
$$ \beta = 8\pi G\sigma \rho *^2
\left(1- cos{\rho_-\over \rho *}\right) ,
\eqno(2.17)$$
$$ \gamma = 8\pi G\tau \rho *^2
\left(1- cos{\rho_-\over \rho *}\right) ,
\eqno(2.18)$$
$$\alpha \rho_+ = \rho * sin{\rho_-\over\rho *}. \eqno(2.19)$$
The remaining metric coefficients are clearly continuous. The only
non zero contorsion terms which enter into (2.10) are
obtained from (2.16) to be
$$\eqalignno{K_{(\hat\phi\hat t)\hat\rho}&=-4\pi G\sigma\rho *
sin{\rho\over\rho *}, K_{(\hat\phi\hat z)\hat\rho}=4\pi
G\tau\rho *
sin{\rho\over\rho *}, &\cr
K_{\hat\phi\hat\phi\hat\rho}&=(8\pi
G)^2 \left( \tau^2-\sigma^2\right)
\rho *^3\left(1- cos{\rho\over
\rho *}\right)
sin{\rho\over\rho *}.&(2.20)\cr}$$
Using (2.19) and(2.20), it can now be verified that the remaining
junction conditions (2.10) are satisfied provided
$\alpha = cos{\rho_-\over\rho *}.$
The mass per unit length is
$$\mu \equiv \int_\Sigma \epsilon\theta^1\wedge\theta^2
={1\over
4G}\left(1-cos{\rho_-\over\rho *}\right) ={1\over 8\pi
G}\int_\Sigma {R^1}_2, \eqno(2.21)$$
where $\Sigma$ is a cross-section of the string (constant $t,z$).
Therefore,
$\alpha = 1-4G\mu$. The angular momentum per unit length due
to
the spin density is
$$J\equiv \int _\Sigma\sigma\theta^1\wedge\theta^2
=2\pi\sigma\rho *^2\left(1-cos{\rho_-\over\rho
*}\right)={1\over 8\pi G}\int_\Sigma Q^0.\eqno(2.22)$$
Hence, from (2.17), $\beta =4GJ$. The angular momentum flux,
which is along the $z-$axis, is
$$F\equiv \int _\Sigma\tau \theta^1\wedge\theta^2
=2\pi\tau\rho *^2\left(1-cos{\rho_-\over\rho
*}\right)={1\over 8\pi G}\int_\Sigma Q^3.\eqno(2.23)$$
Hence, from (2.18), $\gamma =4GF$.

The Sagnac phase shift and the new phase shift obtained earlier
are therefore $\Delta\phi_E = ET^0$, and $\Delta\phi_p = pT^3$,
where $T^0$ and $T^3$ are the fluxes of
$Q^0$ and $Q^3$ through $\Sigma$. These are both topological
phase shifts, analogous to the Aharonov-Bohm effect with the
string playing the role of the solenoid, in that they are invariant
as the curve $\gamma$ is deformed so long as it is outside the
string.

It was recently pointed out to me that the special case of the
above solution corresponding to $\gamma
=0=\tau$ was found by Soleng [27]. If torsion is
absent so that the spin density is zero in the above solution, then
$\beta =0=\gamma$, and the
above solution
reduces to the exact static solution of Einstein's theory found by
Gott
[15] and others [28], whose linearized limit was previously
found by Vilenkin [14].
\vskip .5cm
\noindent{3. INTERACTION OF A QUANTUM COSMIC STRING WITH
A QUANTUM PARTICLE}

Suppose now that the
cosmic string is treated quantum mechanically. Then its
gravitational field also should be treated quantum mechanically.
It is then
possible to form a quantum superposition of the gravitational
fields corresponding to different values of $(\alpha,
\beta,\gamma)$ of the solution obtained above.

It was shown [29]
that the following new physical effect is obtained when the cosmic
string is in a superposition of quantum states corresponding to
different values of $\beta$:
A measurement on a quantum cosmic string that puts it in this
superposition of geometries would change the
intensity of the wave function of a particle in a {\it simply
connected} region near the cosmic string, even though each of the
superposed flat geometries in this region has no effect on the
wave function. This is unlike the
Aharonov-Bohm effect in which the wave function needs to go all
the way around the multiply connected region surrounding the
solenoid in order to be affected by the solenoid.

I shall now treat this effect using the variables $\theta^a$ and
${\omega^a}_b$, and generalize this effect further. Owing to the
translational symmetry along the direction of the string, its
gravitational field is equivalent to that of a point particle in $2+1$
dimensional gravity. Using the latter variables, Witten [30] has
constructed a quantum theory of $2+1$ dimensional gravity which
is finite. The effect which will be treated now is therefore also
obtained in and
provides physical meaning to $2+1$ dimensional quantum gravity.

In the gravitational phase operator (1.2), $\theta^a$ and
${\omega^a}_b$ are now operators owing to the fact that the
gravitational field they represent is quantized. Using (2.3) and
(2.4), (1.2) may be written in terms of $\alpha ,\beta$ and
$\gamma$, which are also operators, as the product of two
commuting exponentials:
$$\eqalignno{\Phi_\gamma &=  \exp\left[ - i\int_{\gamma}
(dt P_0 +\beta d\phi
P_0 + dz P_3 + \gamma d\phi P_3 )\right]\cr
&\times P\exp \left[- i\int_{\gamma}
\left( d\rho P_1 +\alpha \rho d\phi P_2
-\alpha d\phi {M^2}_1\right)
\right]. &(3.1)\cr}$$
 From the end of section 2 it follows that
$$\alpha = 1-4G\hat\mu , \beta = 4G\hat J, \gamma = 4G\hat F,
\eqno(3.2)$$
where $\hat\mu , \hat J$ and $\hat F$ are the quantum
mechanical operators corresponding to the mass, angular
momentum and angular momentum flux per unit length of the
string. The latter operators are assumed to commute with one
another.

Suppose that the quantum state of the cosmic string is initially in
the superposition
$$
|\psi_0\rangle ={1\over\sqrt{2} }
(|\psi_1\rangle  +|\psi_2\rangle ),
\eqno(3.3) $$
where $|\psi_1\rangle $ and $|\psi_2\rangle $ are normalized
eigenstates of
$\alpha ,\beta$ and $\gamma$ with the same eigenvalue for
$\alpha$ and the other eigenvalues being $(\beta_1, \gamma_1)$
and $(\beta_2, \gamma_2)$ respectively. According to (3.2), these
different values of $(\beta ,\gamma )$ correspond to different
eigenstates of $\hat J$ and $\hat F$, respectively, of the fluid that
the string is made of. So, the superposition (3.3) may be obtained
by putting the quantum mechanical particles which constitute this
fluid in the corresponding superposition by, say, letting them
interact with another quantum mechanical system. Suppose also
that a test
particle outside the string is approximately an eigenstate of its
energy $P_0$ and momentum in the $z$-direction $P_3$, with
eigenvalues $E$
and $p$ respectively. This is possible because the last two
operators commute with each other due to the symmetry of the
gravitational field of the string in the $z$-direction.

The test
particle is initially far away from the string in the normalized
state
$|\zeta_0 \rangle$ and is slowly brought towards the string
without
changing $E$ or $p$. Suppose
the interaction of $|\zeta _0\rangle$ with $|\psi_1\rangle $ and
$|\psi_2\rangle $ changes the state of the combined system to
$|\psi_1\rangle |\zeta_1\rangle $ and
$|\psi_2\rangle |\zeta_2\rangle $ respectively. Then by the
linearity of quantum
mechanics, the interaction of $|\zeta _0\rangle$ with
$|\psi_0\rangle $ gives rise to the entangled state for the
combined system
$$
|\chi\rangle ={1\over\sqrt{2}}
(|\psi_1\rangle |\zeta_1\rangle +|\psi_2\rangle
|\zeta_2\rangle ) .
\eqno(3.4) $$

Now a measurement is made on the string and it is found to be in
the superposition
$$ |\psi \rangle =
a|\psi_1\rangle  +b|\psi_2\rangle ,$$
where $|a|^2+|b|^2=1$. The corresponding state of the test particle,
after normalization,
is
$$|\zeta \rangle =\sqrt{2} \langle \psi |\chi\rangle =
a^* |\zeta_1\rangle +b^* |\zeta_2\rangle .
\eqno(3.5)$$
The wave function corresponding to this state is to a good
approximation
$$\zeta({\bf x},t) = a^* \exp\{ -i(\beta_1E+\gamma_1p)(\phi-
\phi_0)\} +
b^* \exp\{ -i(\beta_2E+\gamma_2p)(\phi-\phi_0)\}
\zeta'_0({\bf x},t) ,
\eqno(3.6)$$
where $\zeta'_0({\bf x},t)$ is independent of $\beta$ and
$\gamma$. To obtain (3.6), one may solve the wave equation for
the
interaction of the test particle with the cosmic string in each of the
states $|\psi_1\rangle $ and $|\psi_2\rangle $ and superpose the
two solutions, or one may act on the state of the combined system
by (3.1). Then $\zeta'_0$ is seen to be the result of the
action on $\zeta_0$ of the part of (3.1) that does not depend on
$\beta$ and $\gamma$ and is therefore the same for both of the
superposed states. The constant $\phi_0$ depends on the phase
difference between these states, which in turn depends on the
details of the interaction.

The intensity is
$$\zeta^*\zeta ({\bf x},t) =  \left(
1+2|ab|\cos [ \{ (\beta_1-\beta_2)E+(\gamma_1-
\gamma_2)p\}
(\phi-\phi_0)+\delta ] \right) |\zeta'_0({\bf x},t)|^2 .\eqno(3.7)$$
It follows that the intensity will oscillate as a function of $\phi$.
The number of oscillations per unit angular distance $\phi$ is
$$\nu = {1\over 2\pi}\{(\beta_1-\beta_2)E+(\gamma_1-
\gamma_2)p\} = {2G\over \pi}\{(J_1-J_2)E+(F_1-
F_2)p\},\eqno(3.8)$$
on using (3.2).
This effect may be regarded geometrically as being due to the
difference between two affine connections, which is a tensor field.
This explains why this effect may occur for a wave function that is
in a simply connected region outside the string. Because unlike
each affine connection which has zero curvature,
and can therefore have
physical influence only through its nontrivial holonomy around
the string, the above tensor field may have local influences.

\vskip .5cm
\noindent{4. QUANTUM GENERAL COVARIANCE AND SPACE-TIME
POINTS}

In general, if there is a quantum superposition of gravitational
fields, by a quantum diffeomorphism, or simply a q-
diffeomorphism, I mean performing different diffeomorphisms on
the
superposed gravitational fields. Then the
physical effect described in section 3 may be shown to be
invariant under a particular
q-diffeomorphism performed on the quantized gravitational field
when $\gamma =0$
[29,31]. I postulate that all physical effects are invariant under all
q-diffeomorphisms. This suggests a generalization of the usual
principle of general covariance for the classical gravitational field
to the following {\it principle of quantum general covariance} in
quantum gravity: The laws of physics should be covariant under
q-diffeomorphisms.

On the other hand, the usual principle of general covariance
requires
covariance of the laws of physics under classical diffeomorphisms,
or c-diffeomorphisms. A c-diffeomorphism is a diffeomorphism
that is the same for all the superposed gravitational fields, and is
thus a special case of a q-diffeomorphism. Therefore, the above
principle of quantum general covariance generalizes the usual
general covariance due to Einstein. Under a c-diffeomorphism, a
given space-time point is mapped to the same space-time point
for
all of the geometries corresponding to the superposed
gravitational fields. This is consistent
with regarding the space-time manifold as real, i.e. a four
dimensional ether.

It is instructive in this context to examine Einstein's resolution of
the hole argument [32]: In 1913, Einstein and
Grossmann [33] considered the determination of the gravitational
field inside a hole in some known matter distribution by solving
the gravitational field equations. If these field equations are
generally covariant, then there are an infinite number of solutions
inside the hole, which are isometrically related by
diffeomorphisms. These geometries, which I shall call Einstein
copies, may however be regarded as different representations of
the same objective physical geometry. This follows if a
space-time point inside the hole is defined operationally as the
intersection of the world-lines of two material particles, or
geometrically by the distances along geodesics joining this point to
material points on the boundary of the hole. Under a
c-diffeomorphism, such a point in one Einstein
copy is mapped to a
unique point in another Einstein copy. Both points may then be
regarded as different representations of the same physical
space-time point or an event. So, if we restrict ourselves to
just c-diffeomorphism freedom,
space-time
may be regarded as objective and real.

But the space-time points
associated with each of the superposed gravitational fields,
which are defined above in a c-diffeomorphism invariant manner,
transform
differently under a q-diffeomorphism. This means that in
quantum gravity space-time
points have no invariant meaning. However, protective
observation suggests that quantum states are real [3].
Consequently, the space-time manifold, which appears to be
redundant, may
be discarded, and we may deal directly with the
quantum states of the gravitational field. This is somewhat
analogous to how the prerelativistic ether was discarded because
it did not permit the Lorentz boost symmetries, or if it did then it
was redundant. But then the curve
$\gamma$ in
the gravitational phase operator
(1.2) cannot be meaningfully defined as a curve in space-time.
The
resolution of this difficulty may be expected to lead us to a
quantum theory of
gravity that may be operational and geometrical.

\vskip .5cm
\noindent{ACKNOWLEDGEMENTS}

I thank H. R. Brown, R. Penrose and J. L. Safko for
stimulating discussions. This
research was supported by NSF grant no. PHY-9307708 and ONR
grant no. R\&T 3124141.

\vskip .5cm
\item{[1]}J. Anandan, Foundations of Physics, {\bf 10,} 601
(1980).
\item{[2]}J. Anandan, Foundations of Physics, {\bf 21,} 1265
(1991).
\item{[3]} Y. Aharonov, J. Anandan, and L. Vaidman, Phys. Rev. A
{\bf 47,} 4616 (1993); Y. Aharonov and L. Vaidman, Phys. Lett. A
{\bf 178,} 38
(1993);  J. Anandan, Foundations of Physics
Letters {\bf 6,} 503 (1993).
\item{[4]}J. Ehlers, F. A. E. Pirani, and A. Schild, in {\it Papers in
Honour of J. L. Synge}, edited by L. O'Raifeartaigh (Clarendon
Press,
Oxford 1972).
\item{[5]} J. Anandan in {\it Quantum Theory and Gravitation},
edited by A. R. Marlow (Academic Press, New York 1980), p. 157.
\item{[6]} J. Anandan, Phys. Rev. D {\bf 15,} 1448 (1977).
\item{[7]}J. Anandan, Nuovo Cimento A {\bf 53,} 221
(1979).
\item{[8]} D. Greenberger, Ann. Phys. {\bf 47,} 116 (1968); D.
Greenberger and A. W. Overhauser, Sci. Am. {\bf 242,}, No. 5, 54
(1980).
\item{[9]}S. Kobayashi and K. Nomizu, {\it Foundations of
Differential
Geometry} (John Wiley, New York, 1963).
\item{[10]} For a Yang-Mills field it is convenient to choose the
``radial''
gauge in which $\Lambda (s,t)$ is the identity for all $s,t$. See, for
example, C.
N. Kozameh and E. T. Newman, Phys. Rev. D {\bf 31,} 801 (1985),
Appendix A for this technique. Then, in (1.6), the path ordering
needs
to be done only in $s$. But for affine holonomy this is not
appropriate
because $\theta^a$ are usually linearly independent which
prevents
the ``radial'' gauge being chosen. See also J. A. G. Vickers, Class.
Quantum Grav. {\bf 4,} 1 (1987), who chooses a different set of
paths.
\item{[11]}C. N. Yang, Phys. Rev. Lett. 33 (1974) 445.
\item{[12]} T. W. B. Kibble, J. Phys. A {\bf 9,}1387 (1976); Phys.
Rep.
{\bf 67,} 183 (1980).
\item{[13]} Y. B. Zeldovich, Mon. Not. R. Astron. Soc. {\bf 192,} 663
(1980).
\item{[14]} A. Vilenkin, Phys. Rev. D {\bf 23,} 852 (1981).
\item{[15]} J. R. Gott III, Astrophys. J. {\bf 288,} 422 (1985).
\item{[16]} K. P. Tod, Class. Quantum Grav. {\bf 11,} 1331 (1994).
\item{[17]} D. V. Gal'tsov and P. S. Letelier, Phys. Rev. D {\bf 47,}
4273 (1993). See also P. S. Letelier, Classical and Quantum Gravity
{\bf 12,} 471 (1995).
\item{[18]}S. Deser, R. Jackiw, and G. 't Hooft Ann. Phys. {\bf152,}
220 (1984).
\item{[19]}P. O. Mazur, Phys. Rev. Lett. {\bf 57,} 929 (1986); {\bf
59,} 2380 (1987).
\item{[20]} J. Anandan in {\it Directions in General
Relativity}, Volume 1, Papers in honor of Charles Misner, edited
by
B. L. Hu, M. P. Ryan and C. V. Vishveshwara (Cambridge Univ.
Press, 1993); J. Anandan.
\item{[21]} A. Ashtekar and A. Magnon, J. Math. Phys. {\bf 16,}
342 (1975); J. Anandan, Phys. Rev. D {\bf 24,} 338 (1981).
\item{[22]} J. Anandan, Phys. Lett. A {\bf 195,} 284 (1994).
\item{[23]} D. W. S. Sciama in Recent Developments in General
Relativity (Oxford 1962). p. 415; T. W. B. Kibble, J. Math. Phys. {\bf
2,} 212 (1961).
 \item{[24]} A. Trautman in {\it The Physicist's Conception of
Nature},
edited
by J. Mehra (Reidel, Holland, 1973).
\item{[25]} A. R. Prasanna, Phys. Rev. D {\bf 11,} 2083 (1975); D.
Tsoubelis, Phys. Rev. Lett. {\bf 51,} 2235 (1983).
\item{[26]} W. Arkuszewski, W. Kopczynski, and V. N. Ponomariev,
Commun. Math. Phys. {\bf 45,} 183 (1975).
\item{[27]} H. H. Soleng, J. Gen. Relativ. and Gravit. {\bf 24,} 111
(1992).
\item{[28]}W. A. Hiscock,
Phys. Rev. D {\bf 31,} 3288 (1985); B. Linet, Gen. Relativ. and
Gravit.
{\bf 17,} 1109 (1985).
\item{[29]} J. Anandan, Gen. Relativ. and Gravit. {\bf 26,} 125
(1994).
\item{[30]} E. Witten, Nucl. Phys. B{\bf 311,} 46 (1988).
\item{[31]} Y. Aharonov and J. Anandan, Phys. Lett. A {\bf160,}
493 (1991); J. Anandan, Phys. Lett. A {\bf164,} 369 (1992).
\item{[32]} See for example, R. Torretti, {\it Relativity and
Geometry} (Pergamon Press, Oxford 1983), 5.6.
\item{[33]} A. Einstein and M. Grossmann, Zeitschr. Math. und
Phys. {\bf 62,} 225 (1913).
\bye